\documentclass[12pt]{article}

\usepackage{cite}
\usepackage[pdftex]{graphicx}
\usepackage{amsmath}
\usepackage[margin=1in]{geometry}

\begin{document}

\title{Agent-Based Implementation of Particle Hopping Traffic Model With Stochastic and Queuing Elements}
%
%
%

\author{Camilla~Champion, and Cody~Champion}

\maketitle

\begin{abstract}
Lagging or halted traffic is bothersome. As such, it is desirable to have a model that can begin to determine the efficiency of various traffic standardizations. Our model intended to create a multifaceted realistic simulation of traffic flow while considering several factors. These factors included: passing conventions, e.g., right except to pass (REP) rule, system perturbation caused by insertion of an accident into the system, accessible number of lanes available with the REP, various human factors such as variation of individual maximum speed and likelihood to pass. A succession of models were created from a variation on an existing single-lane traffic model and adding extra dimensionality to the lattice to include multiple lanes, passing conventions, stochastic elements for individuality, and queuing rules to movement algorithms. We found that the REP is an effective means of increasing the critical density that a system can support. Eliminating human factors and thereby automating the system, results in a 160\% increase in the sustainable critical density of the system. The number of lanes increases the critical density of the system, but the maximum efficiency of the speed distribution remains the same. Excluding system automation, the optimal speed distribution for drivers maximal speed was found to be Beta(5,5). Accidents in stable systems can cause small local jams without causing global jams.
\end{abstract}

\section{Introduction}

Traffic jams are an inconvenience. The typical response has been to increase the capacity of the transportation infrastructure. The design of said infrastructure has typically relied on outdated methods of prediction of traffic flow\cite{magnanti}. In recent years the space available for transportation infrastructure expansion has decreased due to population growth, prompting an increased interest in the design and optimization of transport infrastructure\cite{Gillen}. Therefore, a robust and stable model of the dynamics of multiple lanes of traffic is desired. The model herein is adapted from the 2014 Consortium for Mathematic's Mathematical Contest in Modeling problem; the Keep-Right-Except-To-Pass Rule\cite{MCM}. 

Several factors must be considered when formulating these models. Some of these factors are the effects of passing rules, the impact of traffic perturbations caused by accidents, the effects of speed limits, the flow capacity of multiple lane roads, and the impact of stochastic human factors or lack thereof, on traffic systems. An important part of this field of development is understanding the effect of human behaviors on traffic dynamics.

\subsection{Review of Existing Traffic Models} 
There are two predominant types of traffic flow models, macroscopic and microscopic.

Macroscopic methods, such as fluid dynamics, attempt to analyze traffic flow by looking at conditions and variations within the system rather than individual agents\cite{lighthill}. The common variables examined in these types of models are traffic flow, density, speed, and acceleration/deceleration. Fluid dynamics systems can provide a method to predict typical traffic flows rapidly and are therefore often used to predict traffic in real time. However, this is at the cost of predictive accuracy. Additionally, these approaches ignore the variation induced by the individuals within traffic that can produce significant perturbations in the system as a whole. These models can also fail at predicting the dynamics of multilane traffic disturbances caused by individuals\cite{tagkey1971i, tagkey}. The primary issue that hinders the fluid dynamic models of multilane traffic is that the rules that govern individuals are largely ignored\cite{Cremer} and groups of cars are treated as an average of the characteristics of the members of the group also known as a platoon.

The other model type typically used to predict traffic flows is the microscopic type, particularly cellular automata. The concept of this type of model can be traced as far back as 1948 when they were used to study biological systems\cite{neumann}. Over time these models were developed to describe a grid of cells that were in a binary state, either on or off\cite{Wolfram}. Eventually, these models became very well known as the basis for Conway's "Game of Life"\cite{Gardner}. This development was extremely important as it led to Gosper's demonstration that the "Game of Life," and therefore cellular automata, is computationally universal in that it can mimic arbitrary algorithms, a key component in the turning test\cite{Gosper}. 

These model utilized four main rules, each of these will be explored later.

The environment of the model is always a lattice of cells that are spaced in some uniform pattern. These lattices can be 1, 2 or even 3-dimensional. It is typically assumed that the lattices exist in Euclidean space.

Each cell in the lattice can hold any type of information such as a simple binary state or more complex arrays of variables.

Cells can be clustered and can have greater connectivity in more complex lattices, for example, a 1D lattice will have a cell adjacent to up to two neighbors, but in a 2D lattice, the cell can be adjacent to two neighbors in the 1D plane and an additional two in the second direction.

Each cell may be able to change the state of neighboring cells. This allows for both generation and transmission of information in and through the lattice. This information generation is produced when a discrete time step is completed, and all calculations between cells will be performed in parallel.

\subsection{Description of Traffic Assessment}
There are two main ways of assessing the efficiency of traffic: a qualitative view of congestion and a quantitative value of the maximum density that a system can experience before traffic jams are observed.

\subsection{Congestion}
\begin{figure}
	\begin{center}
		\includegraphics[scale=0.4]{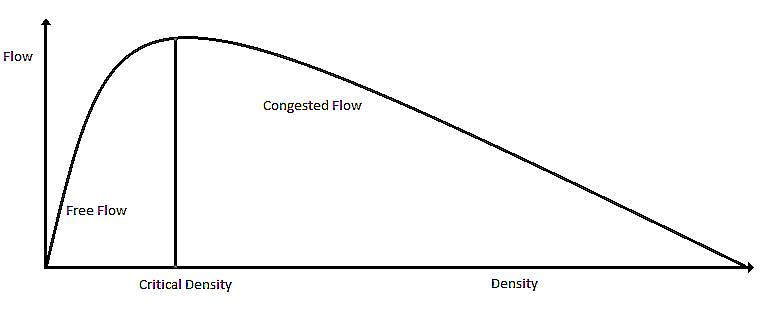}
		\caption{The fundamental diagram of flow and density in traffic systems}
	\end{center}
	\label{Flow}
\end{figure}

Time-space diagrams can illustrate the congestion phenomenon. When the movement of all vehicles is plotted by position and time, it is possible to qualitatively view traffic jams\cite{jam97}. There are two forms of jams, global and local. Global jams will affect all vehicles in the system. All vehicles will decrease in speed or even stop for a period. These types of jams are typically observed in high-density systems typically seven times greater critical density\cite{Kloeden}. Local jams are seen in both high and low-density systems. These occur when one vehicle forces another behind them to slow down before passing. These jams are observable in the space-time diagram as small areas where a small sub-population or platoon of the vehicles are slowed or stopped. This can be seen in the Fig.\ref{Flow}.

\subsubsection{Velocity, Density, and Flow}
Definitions of Velocity (V). This is the number of grid points that a vehicle moves in a single calculation cycle. 

Definition of density (K). This value is defined as the number of vehicles that are on the grid divided by the number of total grid points. The inverse of K is also known as spacing and is used to estimate the safety of the system\cite{Ding}. If the system can maintain high flow at low spacing, then the system will have fewer accidents. 

Therefore, flow is defined as

\begin{equation}
Q =Flow= V K =\frac{Vehicles}{time}\\
\end{equation}
Where
\begin{equation}
V=Velocity=\frac{Grid Points}{time}\\
\end{equation}
and
\begin{equation}
K=Density=\frac{Vehicles}{Grid Area}
\end{equation}

\begin{figure}
	\begin{center}
		\includegraphics[width=\linewidth]{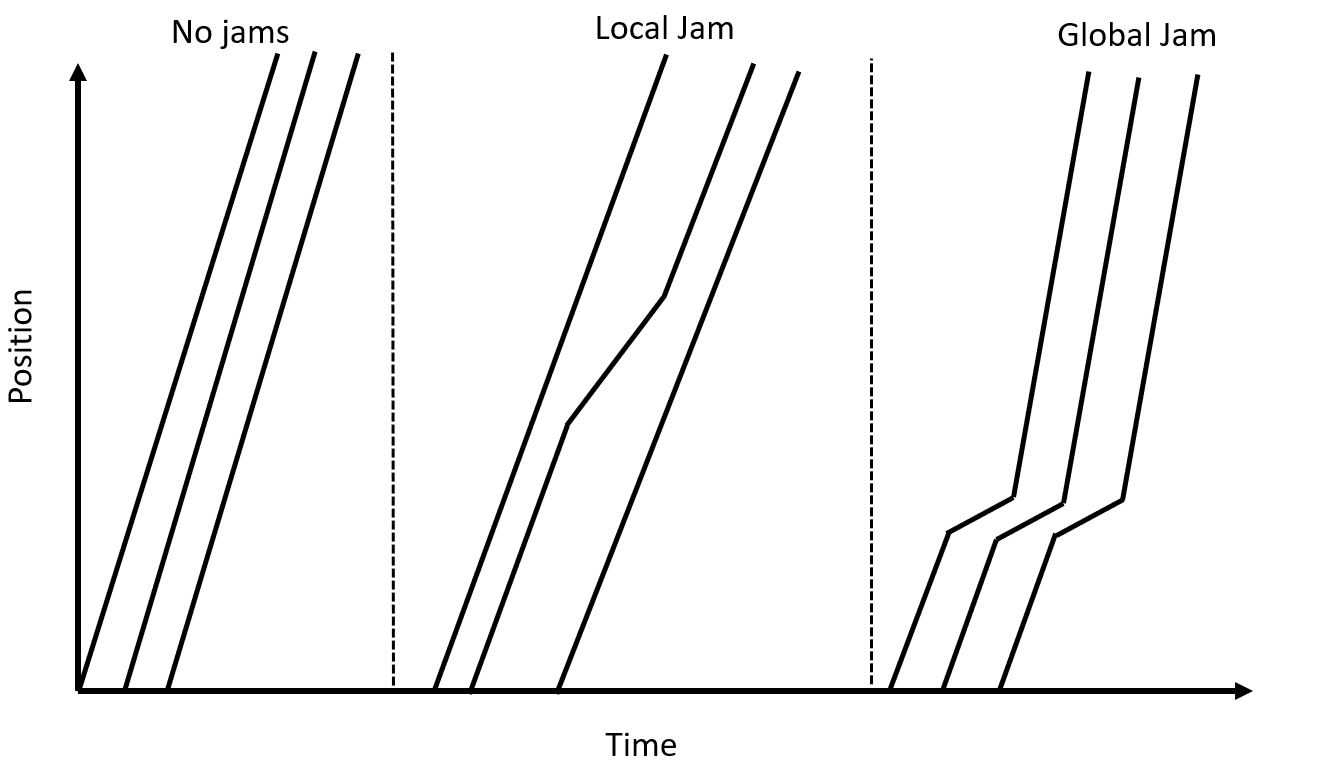}
		\caption{An illustration of a Space-Time Diagram that shows visual representation of traffic jams}
	\end{center}
	\label{jamfig}
\end{figure}

The relationship between flow and density is typically inverse; as density increases the flow decreases. With a graph of flow vs. density, critical density, or $K_{crit}$ can be determined\cite{jam97}. This is the point where the system is first saturated and denotes the maximum density that will not cause global jams. After this point, global jams will occur. A more stable system can accommodate a larger number of cars and higher density. Therefore, a larger critical K value implies a more stable system (See Fig. \ref{jamfig}).

\section{Design}
Because a goal of our model was to be able to determine the effects of having an automated system as opposed to each vehicle having unique characteristics, we chose to develop a microscopic model using elements of cellular automata and more generally, an agent-based model. In our models, each agent, or vehicle, has a set of unique characteristics. In the most simple, the single lane model, this is only its starting position. However, the full model has a set of 9 total characteristics. These are outlined as follows for each vehicle:

\begin{enumerate}
	\item Starting Position: is the position of the vehicle at the start of the time interval
	\item Ending Position: is the position of the vehicle at the end of the time interval
	\item Current Velocity: is the velocity, and thus the distance that the vehicle determines that it should move
	\item Current Lane: is the lane that the vehicle occupies at the start of the time interval
	\item End Lane: is the lane that the vehicle occupies at the end of the time interval
	\item Car Type: Value of 1, or 2 to represent car or truck respectively. This is determined randomly with a researched proportion that is expected to be cars (0.95)\cite{RITA}
	\item Rank of Vehicle: This is only used for the Variable Lanes model and denotes where in the queue of vehicles the vehicle is located
	\item Rudeness Factor: A value from [0,1] denoting the degree of drivers rudeness
	\item Personalized max speed:
\end{enumerate}

These models also use a lattice, or grid, to help track the position of each vehicle similar to the lattices described in the description of microscopic traffic models above. The lattice for these models is defined as follows:
\begin{enumerate}
	\item Has an environment that is composed of a designated number of lanes and gridpoints for a 2-dimensional grid that is the size of the number of lanes by the number of grid points. Vehicles are not allowed to be closer than one gridpoint as shown in the two-lane and eight gridpoints system Fig. \ref{LatticeDist} and will always be a discrete number of gridpoints away.
	
	\begin{figure}
		\begin{center}
			\includegraphics[width=\linewidth]{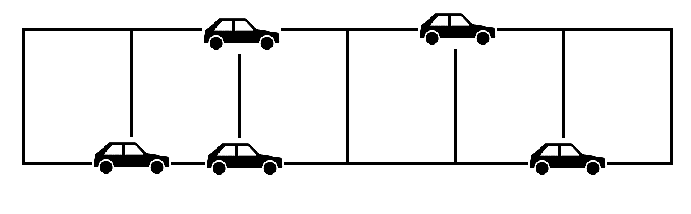}
			\caption{Each vehicle in the lattice can be no closer than 1 gridpoint in either dimension and must be a discrete number of gridpoints away}
		\end{center}
		\label{LatticeDist}
	\end{figure}
	
	\item Each gridpoint holds the vehicle type that is there. In models where there are no trucks, i.e., the single lane system this information is binary. When accidents are introduced, the accident location is also listed as a value in the grid, and no vehicles are allowed to get within one gridpoint of the accident. Trucks occupy two adjacent gridpoints in the same lane and are marked as a two in the front position.
	\item This Lattice has a neighborhood that is two dimensional. Thus, depending on the number of lanes, a vehicle can have up to 4 immediate neighbors
	\item Each vehicle uses the grid to determine their closest neighbors and where they can move.
\end{enumerate}

We steadily built up our models from a single lane system with no differences between agents to a variable lane system able to include or exclude stochastic elements of human behavior in order to simulate human or automated traffic, introduce a halt in traffic that is positioned at the scene of an accident from an accident and determine the effects, determine the $K_{crit}$ values of different velocity distributions for varying initial densities, queuing impacts and determine the efficiency of the keep-right-except-to-pass rule.

\subsection{Universal Assumptions}
All of the created models were designed with the following assumptions:
\begin{enumerate}
	\item Velocities of all vehicles in the system are discretely positive throughout the models (i.e. speed). 
	\item Vehicles will respond to stimuli from forward and backward directions 
	\item Lane changes are instantaneous
	\item After initial vehicle assignment, no vehicles will enter or re-enter the system
	\item A vehicle's response to stimuli can be quantified\cite{jam97}
	\item All vehicles will maintain a set distance between others at the expense of speed
	\item All vehicles will accelerate and decelerate at a constant rate
	\item As vehicle spacing decreases, the chance of accidents will increase
	\item A Vehicle's behavior is consistent
\end{enumerate}

\subsection{Universal Variables}
and the following variables throughout:
\begin{enumerate}
	\item $numlanes$: The number of lanes allowed for the simulation, a constant in each simulation. This is also the number of rows in the lattice.
	\item $density$: Value between 0 and 1, this value is variable when building fundamental diagrams and a constant when building space-time plots.
	\item $gridpts$: Number of columns in the lattice
	\item $numcars$ = round(density*gridpts*numlanes): The number of vehicles allowed for the simulation. When density varies, this also varies.
	\item $vmax$ The maximum velocity that each car is allowed to move. This is varied between simulations sometimes using scaled beta distributions with $\alpha+\beta=10$ and sometimes constant for every vehicle.
	\item $badDriverp$: Constant probability throughout each simulation of a driver randomly slowing down
	\item $passfactor$: The probability that a driver will choose to pass when they are allowed to do so
\end{enumerate}

Though this was not fully built in every model this was the agent matrix:

\subsection{Preliminary Models}

\begin{itemize}
	\item{The Single Lane Model:} This model was created to resemble rule 184\cite{Fuks}. As such the model only had one lane of traffic. Cars, which were all one size, taking up one grid point, looked forward to determine where the closest car in front of them was. 
	\item{The Two-Lane Models:} The goal of these models was to emulate a two-lane highway in which the passing convention could be changed. Therefore, two distinct models were created, one was asymmetrical in which cars must only pass on the left and never the right. The other which is referred to as symmetrical allows for cars to pass in either lane without regard to different sides of the road.
\end{itemize}

\section{Important Results}
\subsection{Fundamental Diagrams}
The critical densities were calculated in systems where human factors were present, and speed of individuals was assigned via beta distribution. Both the asymmetric and symmetric two-lane models were used. The critical densities of uniform maximum speed were calculated (asymmetric 0.12, and the symmetric was 0.08). 

\begin{figure}
	\centering
	\includegraphics[width=\linewidth]{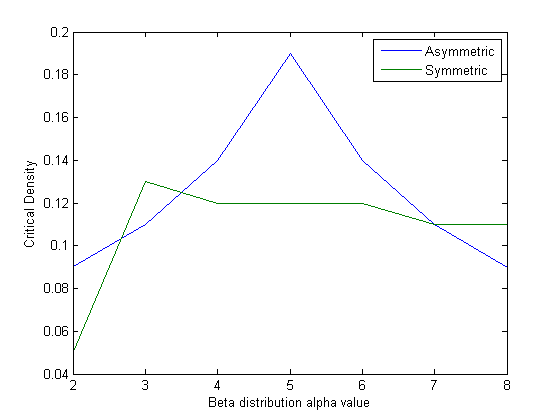}
	\caption{The Critical Densities of systems that had agents maximum speed beta distributed}
\end{figure}

\begin{figure}
	\centering
	\includegraphics[width=\linewidth]{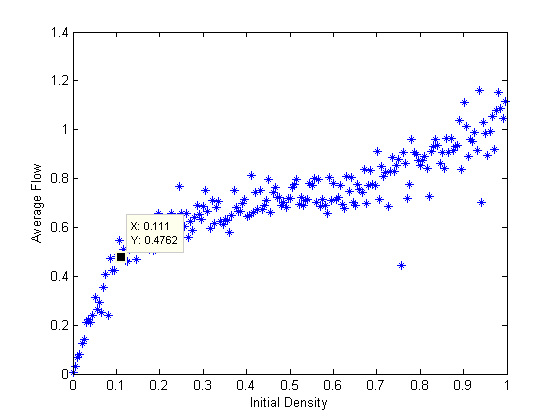}
	\caption{Critical density of passing systems by maximum, as determined by beta distribution.}
	\label{repF}
\end{figure}

\begin{figure}
	\centering
	\includegraphics[width=\linewidth]{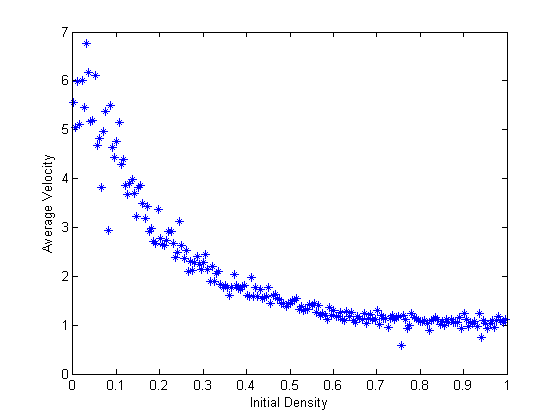}
	\caption{Average velocity and initial density of an asymmetric system with a speed distribution of Beta(7,3)}
	\label{repS}
\end{figure}

A representative plot that was used to find this information is shown in Fig. \ref{repF} and \ref{repS}

\subsection{Discussion of Jams}
\begin{figure}
	\centering
	\includegraphics[width=\linewidth]{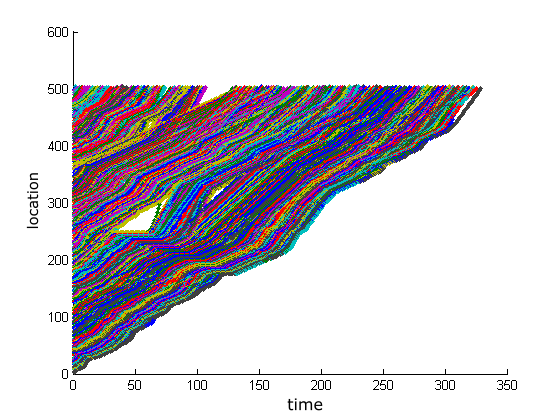}
	\caption{Space Time plot of a Asymmetric system that had a Beta(5,5) speed distribution and an accident. The accident can be seen as a local jam.}
	\label{acc}
\end{figure}

The effects of the changes in the systems were also examined by space-time plots. Accidents were also introduced into the systems as a perturbation. As seen in Fig. \ref{acc} the accident did perturb the system but once the accident cleared the system returned to normal flow. 

\begin{figure}
	\centering
	\includegraphics[width=\linewidth]{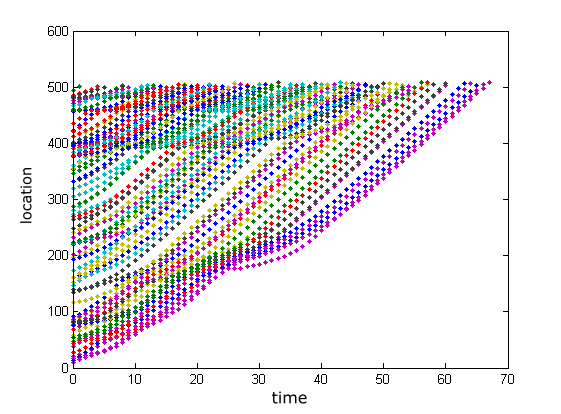}
	\caption{Space Time diagram of symmetrical system at critical density with a uniform speed distribution, this is representative of low capacity systems}
	\label{symtime}
\end{figure}

\begin{figure}
	\centering
	\includegraphics[width=\linewidth]{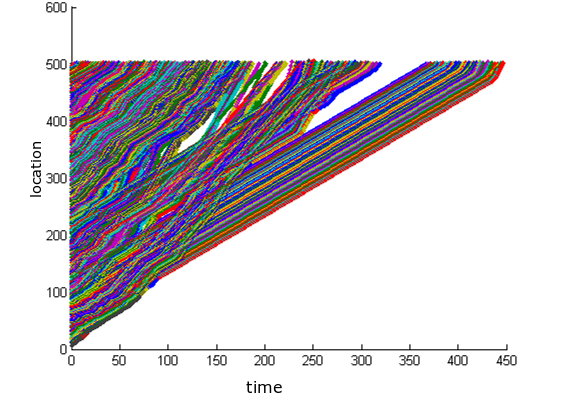}
	\caption{Space Time diagram of an asymmetric system at twice critical density with speed distribution of Beta(5,5). This is representative of a high capacity system.}
	\label{highcap}
\end{figure}

Both symmetric and asymmetric models were examined, and the occurrence of global jams was noted. Shown in Fig. \ref{symtime} a low critical density system was shown, the symmetrical model at its maximum critical density. Note the occurrence of multiple global jams. In Fig. \ref{highcap} a high capacity system is shown, the reason that this system is capable of high-density flow is that jams are local and not global. An accident caused global jams in the symmetric system and only local jams in the asymmetric.

\subsection{Queuing Effects}

\begin{figure}
	\centering
	\includegraphics[width=\linewidth]{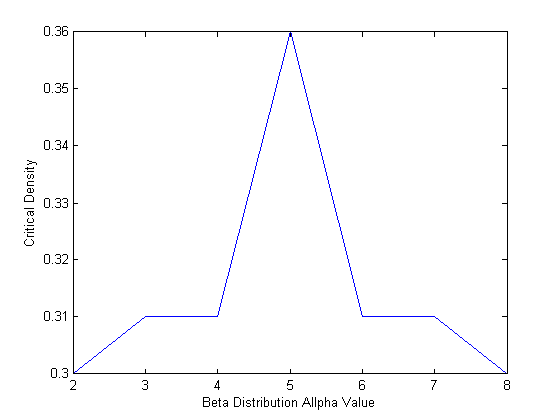}
	\caption{The density and speed distribution plot of a 4 lane system}
	\label{4lane}
\end{figure}

The effects of systems with more than two lanes with queuing effects were also investigated. These results are shown in Fig. \ref{4lane}.

The critical density is the same in the optimal system when four lanes are used instead of 2, but the baseline Critical density has increased when queuing effects are considered.    

\section{Conclusion}

The REP is an effective means of increasing the critical density that a system can support. If human factors are eliminated, i.e. the system is automated, the critical density that the system can support will also be increased by a factor of approximately 160 percent. The number of lanes in a highway will increase the critical density of the system, but the relationship between the critical density and speed distribution of drivers will remain proportional. Without automation, the optimal speed distribution for divers maximum speed is Beta(5,5). Accidents in stable systems can cause small local jams without causing global jams. All code is available from the associated Github repository\cite{github}.

\subsection{Safety}
By examining the space-time plots, we can see if the system is producing a large amount of stop and go traffic. The least change in speed in the system the fewer accidents occur \cite{Kloeden}. The systems that had high critical densities showed least stop and go traffic as expected because local jams will cause larger global jams. Therefore the proposed optimal systems are not only faster but safer as well.

\subsection{Strength of Models}
Due to the microscopic nature of the model we were able to measure the speed at every calculation cycle and then use these values to calculate the global speed of every vehicle that is still on the grid. Other models lack this capability and must use alternative methods of measuring speed. Our variables include stochasticity which mimics real-world variations to a higher degree than fluid dynamics models. Additionally, the addition of queuing, as would be representative of movement in actual traffic distinguishes our model from a cellular automata model and provides more relevant $k_{crit}$ values.

\subsection{Future work}
The improvements on this model could be made are increasing the number of variables that are assigned to each agent. Some possible variables are: weather (would affect all vehicles on the road by lowering the maximum speed and also by changing the reaction times of drivers),  mental load (consider the human factor of distraction, for example, a driver would have a slower reaction time if they are surrounded by other vehicles since their attention will be divided), other types of vehicles (motorcycles and extra wide load trucks would all have different rules associated with them, motorcycles could ignore gaps and weave through traffic, or extra wide load trucks would take up two lanes).

\section*{Acknowledgment}
Cody Champion was supported by the National Science Foundation Graduate Research Fellowship Program under Grant No. 1144468. The content is solely the responsibility of the authors and does not necessarily represent the official views of the National Science Foundation.

\bibliographystyle{IEEEtran}

\bibliography{mathbib}

\end{document}